\documentclass[twocolumn]{aastex631}

\usepackage{graphicx}
\usepackage{amsmath}
\usepackage{color}
\usepackage{comment}
\usepackage{units}
\usepackage[normalem]{ulem}

\newcommand{\SPA}{School of Physics and Astronomy, Monash University, Clayton VIC 3800, Australia}
\newcommand{\OzGravMonash}{OzGrav: The ARC Centre of Excellence for Gravitational Wave Discovery, Clayton VIC 3800, Australia}

\begin{document}

\title{A Gaussian process framework for testing general relativity with gravitational waves}

\author{Lachlan Passenger}
\email{lachlan.passenger1@monash.edu}
\affiliation{\SPA}
\affiliation{\OzGravMonash}

\author{Shun Yin Cheung}
\affiliation{\SPA}
\affiliation{\OzGravMonash}

\author{Nir Guttman}
\affiliation{\SPA}
\affiliation{\OzGravMonash}

\author{Nikhil Kannachel}
\affiliation{\SPA}
\affiliation{\OzGravMonash}

\author{Paul D. Lasky}
\affiliation{\SPA}
\affiliation{\OzGravMonash}

\author{Eric Thrane}
\affiliation{\SPA}
\affiliation{\OzGravMonash}

\begin{abstract}
Gravitational-wave astronomy provides a promising avenue for the discovery of new physics beyond general relativity as it probes extreme curvature and ultra-relativistic dynamics.
However, in the absence of a compelling alternative to general relativity, it is difficult to carry out an analysis that allows for a wide range of deviations.
To that end, we introduce a Gaussian process framework to search for deviations from general relativity in gravitational-wave signals from binary black hole mergers with minimal assumptions.
We employ a kernel that enforces our prior beliefs that---if gravitational waveforms deviate from the predictions of general relativity---the deviation is likely to be localised in time near the merger with some characteristic frequency.
We demonstrate this formalism with simulated data and apply it to events from Gravitational-Wave Transient Catalog~3. 
We find no evidence for a deviation from general relativity. 
We limit the fractional deviation in gravitational-wave strain to as low as $7\%$ (90\% credibility) of the strain of GW190701\_203306. 
\end{abstract}


\section{Introduction}
The direct detection of gravitational waves (GWs) by the LIGO \citep[][]{theligoscientificcollaborationAdvancedLIGO2015}, Virgo \citep[][]{acerneseAdvancedVirgo2nd2015a}, and KAGRA \citep[][]{akutsuKAGRAGenerationInterferometric2019} collaborations has provided a unique opportunity to test general relativity (GR) in the strong-field regime. 
As the exact form a deviation from GR may take is unclear, a variety of tests have been proposed.

Inspiral-merger-ringdown consistency tests check if the early and late stages of binary evolution are consistent \citep[see, e.g.,][]{hughesGoldenBinaryGravitationalWave2005, ghoshTestingGeneralRelativity2017, abbottTestsGeneralRelativity2016, testinggrgwtc1, testinggrgwtc2, testinggrgwtc3}. Parameterised post-Newtonian tests are performed by introducing deviations in the post-Newtonian waveform coefficients predicted for GR \citep[see, e.g.,][]{abbottTestsGeneralRelativity2016, testinggrgwtc1, testinggrgwtc2, testinggrgwtc3}.
In residual strain tests, several flexible models have been used to search for unmodelled, coherent excess power in gravitational-wave residual data (after the best-fit GR template has been subtracted). These include spline models, which consist of smoothly-connected piecewise polynomials \citep[][]{edelmanConstrainingUnmodeledPhysics2021}, and algorithms such as \textsc{BayesWave} \citep[][see e.g., \citet{ abbottTestsGeneralRelativity2016, testinggrgwtc1, testinggrgwtc2, testinggrgwtc3}]{cornishBayeswaveBayesianInference2015}, which models signals as a sum of flexible wavelets and allows the number of wavelets to vary during sampling.

Gaussian process regression is a flexible modeling technique that assumes the data points (in our case, strain time series data) are drawn from a multivariate Gaussian distribution. 
A kernel is employed to determine which combinations of data points are most likely.
Existing applications of Gaussian process regression in gravitational-wave astronomy include: glitch mitigation \citep[][]{ashtonGaussianProcessesGlitchrobust2023}; waveform approximant error mitigation \citep[][]{liuImprovingScalabilityGaussianprocess2023};
estimating probability density functions of gravitational-wave event posteriors \citep[][]{demilioDensityEstimationGaussian2021}; and tests of GR focused on gravitational-wave propagation effects~\citep[][]{belgacemGaussianProcessesReconstruction2020, canas-herreraLearningHowSurf2021}.
Here, we propose a Gaussian process formalism to search for deviations from GR in gravitational-wave strain data.
This is a companion paper to \citet{cheung_search_2025}, which develops a similar Gaussian process framework to search for extra polarisations in gravitational-wave strain data.

We structure this paper as follows. In Section \ref{formalism}, we describe our Gaussian process framework.
In Section \ref{demonstration} we demonstrate this framework with simulated data. In Section \ref{residual_data}, we apply our method to the 60 binary black hole gravitational-wave events from the third Gravitational-Wave Transient Catalogue \citep[GWTC-3,][]{ligoscientificcollaborationGWTC3CompactBinary2023} detected by both the LIGO Hanford (H1) and Livingston (L1) observatories. We find no evidence for deviations from GR, and limit the fractional deviation in gravitational-wave strain to as low as $7\%$ (90\% credibility) of the strain of GW190701\_203306.

\section{Formalism}\label{formalism}
\subsection{The Gaussian process likelihood}\label{the_GP_likelihood}
Our strain data $h$ consists of three components: a signal predicted by GR $s_\text{GR}$, instrumental noise $n$, and a signal from physics beyond GR $\delta s$:\footnote{In practice, $\delta s$ also contains unmodelled effects such as noise and waveform systematics---see \citet[][]{guptaPossibleCausesFalse2024} for a recent review. We discuss this further in Section \ref{discussionconclusions}.}
\begin{align}
    h = s_\text{GR} + n + \delta s.
\end{align}  
The signal predicted by GR $s_\text{GR}$ is calculated assuming a waveform approximant; these depend on binary parameters, denoted $\theta$. In this work we employ the \textsc{IMRPhenomPv2} \citep[][]{schmidtModelsGravitationalWaveforms2012} waveform approximant, though we note that this choice is not unique to this framework, and encourage the use of other waveform approximants with different assumptions of the underlying physics of gravitational wave sources.

We take the noise to be Gaussian.
In the frequency domain, the covariance between the noise in different frequency bins is described approximately by a diagonal matrix:
\begin{align}\label{eq:N}
    \textbf{\textit{N}}_{ij} \equiv & \langle n(f_i)^* \, n(f_j) \rangle \\
    = & \frac{1}{4 \Delta f} P(f_j), \delta_{ij} , 
\end{align} 
where $P(f_j)$ is the single-sided noise power spectral density \citep[]{intro},\footnote{In practice, the data are windowed with some function $w(t)$, which reduces the noise power spectral density by a factor of $\overline{w^2}$.} $\Delta f$ is the frequency-bin width and $\delta_{ij}$ is a Kronecker delta. Here and throughout, repeated indices do not imply summation.

We model new physics as a Gaussian process so that $\delta s$ is completely described by a (non-diagonal) ``signal covariance matrix'':
\begin{align}\label{eq:S}
    \textbf{\textit{S}}_{ij} \equiv \langle \delta s^*(f_i) \, \delta s(f_j) \rangle .
\end{align}
Given this assumption, deviations from GR are described probabilistically rather than with parameterised waveforms \citep[e.g.,][]{Abbott2016,Abbott2021}.
The form of $\textbf{\textit{S}}$ is determined by our prior beliefs about deviations from GR.
In general, $\textbf{\textit{S}}$ may depend on parameters that we denote $\Lambda$.

\subsection{Kernel design}\label{kernel_design}
To derive an expression for $\textbf{\textit{S}}$, we first design a kernel $\textbf{\textit{K}}$, which describes the covariance between $\delta s (t)$ at different times.\footnote{Here, we implicitly assume that the deviation from GR produces a GW that is $+,\times$ polarised. The more general case, where the deviation appears as a non-standard polarisation mode, is discussed in the companion paper \citet{cheung_search_2025}.}
In choosing our kernel, we adopt the following prior beliefs about $\delta s (t)$:
\begin{itemize}
    \item The deviation is characterised by some characteristic frequency $f_0$, which is similar to the merger frequency.
    \item The deviation is localised in time so that $\delta s (t)$ goes to zero before and after the merger on a time scale comparable to the inverse characteristic frequency $1/f_0$
    \item The time series $\delta s (t)$ is not necessarily symmetric in time.
\end{itemize}

\begin{figure}
    \centering
    \includegraphics[width=0.49\textwidth]{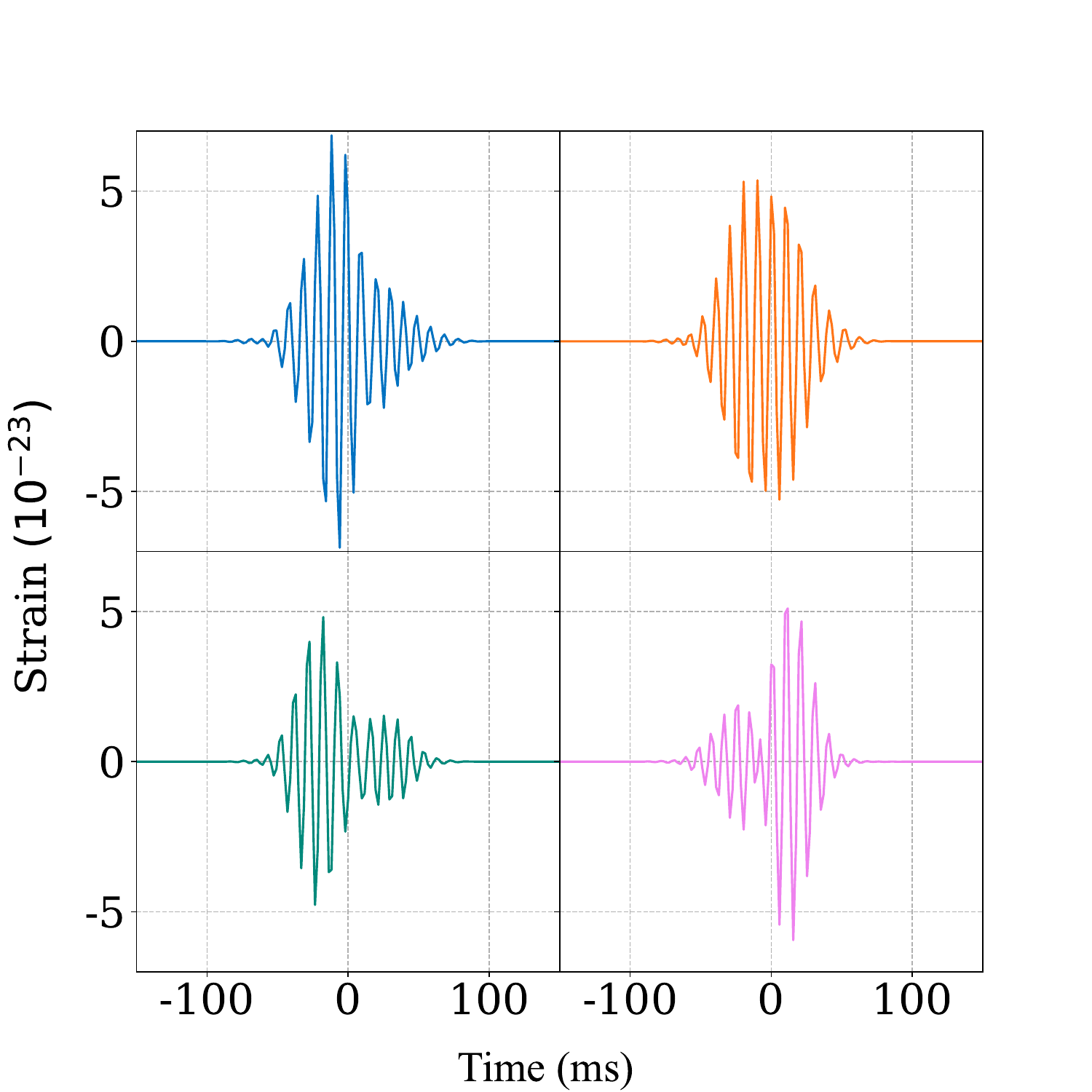}
    \caption{
    Example deviations from GR drawn from our kernel (Eq. \ref{eq:Kt}), with scale factor $k_0=10^{-45}$, characteristic frequency $f_0 = \unit[100]{Hz}$, width $w=2.0$, and length $l = 2.5$. 
    }
    \label{fig:kdraws}
\end{figure}

In order to enforce these prior beliefs, we introduce the time-domain kernel matrix
\begin{align}\label{eq:Kt}
    \textbf{\textit{K}}_{ij} = & K(t_i, t_j) \nonumber\\
    = &
    k_0 e^{-f_0^2 (t_i^2+t_j^2) / 2 w^2}
    \cos\left(2\pi f_0 \tau_{ij}\right)
    e^{-f_0^2 \tau_{ij}^2 / 2l^2} ,
\end{align}
where
\begin{align}
    \tau_{ij} \equiv | t_i - t_j |, 
\end{align}
is the absolute value of the difference of two sample times $t_i \text{ and } t_j$.
In Fig.~\ref{fig:kdraws}, we plot draws of $\delta s(t)$ from the kernel (Eq. \ref{eq:Kt}).
These random draws showcase that this kernel produces the features consistent with our prior beliefs.
The kernel is composed of several commonly used kernel functions, which together enforce our prior beliefs:
\begin{itemize}
    \item The Gaussian kernel $e^{-f_0^2 (t_i^2+t_j^2) / 2 w^2}$ allows for $\delta s (t)$ to be localised in time.
    \item The cosine kernel $\cos\left(2\pi f_0 \tau_{ij}\right)$ allows the amplitude of $\delta s (t)$ to oscillate.
    \item The radial basis function  kernel $e^{-f_0^2 \tau_{ij}^2 / 2l^2}$ allows for some degree of stochasticity in $\delta s (t)$.
\end{itemize}

These kernel functions depend on several parameters:
\begin{itemize}
    \item The scale factor $k_0$ determines the overall amplitude of $\delta s (t)$. The typical strain amplitude for a deviation scales like $k_0^{1/2}$.
    \item The width $w$ determines the typical duration of the signal $\delta s (t)$.
    \item The characteristic frequency $f_0$ sets the time scale for typical oscillations in $\delta s (t)$, and its inverse sets the time scale for the deviations to go to zero.
    \item The coherence length $l$ determines the typical number of cycles over which the frequency of $\delta s (t)$ is coherent; large values of $l$ produce more sinusoidal waveforms whereas small values produce more stochastic waveforms.
\end{itemize}
This choice of kernel reflects our own prior beliefs; we encourage the development of other kernels.

Since the noise in gravitational-wave observatories is described in the frequency domain, the kernel from Eq.~\ref{eq:Kt} must be transformed before it can be incorporated with the detector noise in the likelihood function. 
The frequency-domain kernel is 
\begin{align}\label{frequencycutoff} 
    \textbf{\textit{K}}(f) = \textbf{\textit{H}}^\dagger \textbf{\textit{U}} \textbf{\textit{K}}(t) \textbf{\textit{U}}^\dagger \textbf{\textit{H}} , 
\end{align}
where $\textbf{\textit{U}}$ is the unitary discrete Fourier transform matrix and $\textbf{\textit{H}}$ is a projection operator that removes all frequencies outside of the observing band. 
For this study, we take the band to be $\unit[50-512]{Hz}$.
The minimum frequency of $\unit[50]{Hz}$ is motivated by the presence of non-stationary noise in the $\unit[20-50]{Hz}$ range, e.g., due to scattered laser light inside the detector \citep[see, e.g.][]{accadiaNoiseScatteredLight2010, soniReducingScatteredLight2020, tolleyArchEnemyRemovingScatteredlight2023}. A recent analysis by \citet{shun_memory} highlighted the impact of such noise on an analysis of gravitational-wave memory signals in several gravitational-wave events. Accounting for this noise would likely require a more complicated noise model, which we leave for future work.

In choosing the high-frequency cut-off of $\unit[512]{Hz}$, we assume that a typical deviation from GR is likely to have a characteristic frequency $f_0$ somewhat similar to the ringdown frequency of the binary merger it is associated with. Specifically, we assume that the characteristic frequency of a deviation from GR should not be much more than twice the ringdown frequency. For a GW150914 like event, where the ringdown frequency is $\sim \unit[260]{Hz}$, this gives an upper bound on the frequency of a deviation from GR of $\sim \unit[520]{Hz}$, which we round to the nearest power of two as $\unit[512]{Hz}$. 
Low-mass events produce ringdowns with frequencies higher than GW150914-like events.
However, these ringdowns are buried under shot noise above $\approx\unit[500]{Hz}$, which justifies focusing for now on high-mass events in the $\unit[50-512]{Hz}$ band.

\subsection{Detector response}\label{cov_matrix_construction}
Given our assumption that the deviation from GR consists of $+$ and $\times$ polarised GWs, the strain measured in detector $\mu$ is
\begin{align}
    \delta s_\mu = F_{\mu,+} \, \delta s_+ +
    F_{\mu,\times} \, \delta s_\times .
\end{align}
Here, $F_{+,\times}$ are the antenna response functions, which depend on the sky location $(\text{ra}, \text{dec})$ and polarization angle $({\psi})$ \citep[][]{Nishizawa}.
The signal in detector $\nu$ is related to the signal in detector $\mu$ by a phase shift
\begin{align}
    \delta s_{+,\times}^\nu (f_i) =  
    e^{2\pi i f_i \tau_{\mu\nu}} \, 
    \delta s_{+,\times}^\mu (f_i) ,
\end{align}
where $\tau_{\mu\nu}$ is the time delay between detectors $\mu$ and $\nu$, which implicitly depends on the sky location of the source. 
We assume that the $+$ and $\times$ modes are uncorrelated,\footnote{One is free to adopt other assumptions, for example, that the source is linearly polarised.} such that 
\begin{align}
    \langle \delta s_+^*(f_i) \, \delta s_\times(f_j) \rangle =
    \langle \delta s_\times^*(f_i) \, \delta s_+(f_j) \rangle = 0 .
\end{align}
We assume that the $+$ and $\times$ modes behave the same way on average, such that
\begin{align}
    \langle \delta s_+^*(f_i) \, \delta s_+(f_j) \rangle = &
    \langle \delta s_\times^*(f_i) \,  \delta s_\times(f_j) \rangle \nonumber\\
    \equiv &  \textbf{\textit{K}}_{ij} .
\end{align}
We assume each realisation of $\delta s_+(f_i)$ and $\delta s_\times(f_j)$ are independent draws from a normal distribution with covariance $\textbf{\textit{K}}_{ij}$, which depends on parameters $\Lambda \equiv \{k_0, w, f_0, l\}$.
Given our assumptions, the signal covariance matrix $\textbf{\textit{S}}$ (defined in Eq.~\ref{eq:S}) can be written in terms of the kernel $\textbf{\textit{K}}$ and the antenna factors for two detectors as a block matrix:\footnote{We derive the form of $\textbf{\textit{S}}$ for $>2$ detectors in Appendix \ref{different_detectors}.}
\begin{widetext}
\begin{align}\label{eq:Sij}
    \textbf{\textit{S}}_{ij} = & 
    \left(
    \begin{matrix}
    \textbf{\textit{S}}_{ij}^{\mu\mu} & \textbf{\textit{S}}_{ij}^{\mu\nu} \\
    \textbf{\textit{S}}_{ij}^{\nu\mu} & \textbf{\textit{S}}_{ij}^{\nu\nu}
    \end{matrix}
    \right)
    \nonumber\\
    = & 
    \left(
    \begin{matrix}
        (F_{\mu, +}^2 + F_{\mu,\times}^2) \textbf{\textit{K}}_{ij} &  e^{2\pi i f_j \tau_{\mu\nu}} (F_{\mu,+}F_{\nu,+} + F_{\mu,\times}F_{\nu,\times}) \textbf{\textit{K}}_{ij} \\
         e^{-2\pi i f_j \tau_{\mu\nu}} (F_{\mu,+}F_{\nu,+} + F_{\mu,\times}F_{\nu,\times}) \textbf{\textit{K}}_{ji}^*  & (F_{\nu,+}^2 + F_{\nu,\times}^2) \textbf{\textit{K}}_{ij}
    \end{matrix}
    \right) .
\end{align}
\end{widetext}

\subsection{Building the likelihood}
Next, we subtract the GR waveform from the data to work with the likelihood for the residual data
\begin{align}
    \delta h \equiv & h - s_\text{GR} \nonumber\\
    = & F_+ \delta s_+ + F_\times \delta s_\times + n .
\end{align}
Since $s_\text{GR}$ depends on the binary parameters $\theta$, $\delta h$ also depends on $\theta$.
For simplicity, we assume below that the maximum-likelihood estimate of $s_\text{GR}$ perfectly subtracts the GR signal from the data.
However, one can in principle marginalise over uncertainty in $\theta$ with importance sampling. We demonstrate this marginalisation in Section \ref{discussionconclusions}. We find that this has a minimal effect on our confidence in detections of deviations from GR. Therefore, we will proceed using the max-likelihood estimate of $s_\text{GR}$.

The likelihood function for complex strain data $\delta h$ is given by \citep[see, e.g.,][]{veitchParameterEstimationCompact2015}

\begin{align}\label{eq:L}
    {\cal L}(\delta h | \Lambda) = &
    \frac{1}{\pi^n \det \textbf{\textit{C}}(\Lambda)} \exp 
    \left(- \delta h^\dagger \textbf{\textit{C}}^{-1} (\Lambda) \delta h \right),
\end{align}

where $n$ is the number of frequency bins and  $\textbf{\textit{C}}(\Lambda)$ is the total covariance matrix
\begin{align}\label{eq:C}
    \textbf{\textit{C}}(\Lambda) = \textbf{\textit{S}}(\Lambda) + \textbf{\textit{N}} .
\end{align}
The signal covariance matrix adds simply with the noise matrix since the signal is not covariant with the noise.

\section{Demonstration}\label{demonstration}
\subsection{Tests on simulated signals}\label{simulateddeviations}
We first analyse a simulated signal drawn from the signal covariance matrix $\textbf{\textit{S}}$ (Eq. \ref{eq:Sij}) with hyper-parameter values $k_0=2.0\times 10^{-43}$, $f_0 = \unit[100]{Hz}$, $w=2.0$, and $l = 2.5$. 
We add this signal to a randomly selected $\unit[2]{s}$ long data segment (beginning at GPS = $\unit[1267702309]{s}$) from the LIGO-Virgo-KAGRA collaboration's third observing run (O3), during which both the LIGO H1 and L1 observatories were in observing mode and no GW signal is evident. 
The injection has an optimal network signal-to-noise ratio (SNR) of $9.4$. 
We use gravitational-wave strain data from the Gravitational Wave Open Science Centre \cite[GWOSC; ][]{abbottOpenDataThird2023}, and perform Gaussian process regression using the nested sampler \textsc{dynesty} \citep[][]{speagleDynestyDynamicNested2020} as implemented in \textsc{Bilby} \citep[][]{ashtonBilbyUserfriendlyBayesian2019,bilby_gwtc1}. 
We construct the noise power spectral density (PSD, Eq. \ref{eq:N}) using the $\unit[64]{s}$ of data prior to the GPS trigger time. We split the data into overlapping segments and use the median to estimate the PSD.

We choose priors on the Gaussian process hyper-parameters as follows: $f_0$ is distributed uniformly on the interval $[\unit[50]{Hz},\unit[560]{Hz}]$, as we expect this range to capture a deviation from GR arising in a GW150914-like event, as discussed in Subsection \ref{kernel_design}. On average, signals with characteristic frequencies above $\unit[560]{Hz}$, with other parameters drawn from their respective priors, have more than $90\%$ of their SNR above the maximum observing frequency of $\unit[512]{Hz}$.
Our priors on $w$ and $l$ are log-uniformly distributed on the interval $[0.1,5]$. The lower bound of these priors is chosen to include signals in which approximately half a cycle is visible in the time domain strain of typical signals drawn from the kernel. The upper bound on these parameters is chosen so that a maximum of approximately 10 to 20 cycles are visible.
Finally, $k_0$ is log-uniform distributed over the interval $[10^{-46}, 10^{-41}]$. 
The minimum bound is chosen to correspond to a signal that is indistinguishable from zero; the average optimal network signal-to-noise ratio for a signal with $k_0 = 10^{-46}$, with other parameters drawn from their associated priors, is approximately $0.10$.
The maximum bound is chosen to correspond to an unambiguous detection; the average optimal network signal-to-noise ratio for a signal with $k_0 = 10^{-41}$, with other parameters drawn from their associated priors, is approximately $30$.\footnote{These numbers are obtained by analysing a \unit[2]{s} long segment of typical noise beginning at GPS trigger time $\unit[1267702309]{s}$.} We choose a log-uniform distributed prior for $w$, $l$ and $k_0$ as we do not know the typical orders of magnitude of the length scale, or amplitude, a deviation from GR will manifest as.

\begin{figure*}
    \centering
    \includegraphics[width=12cm]{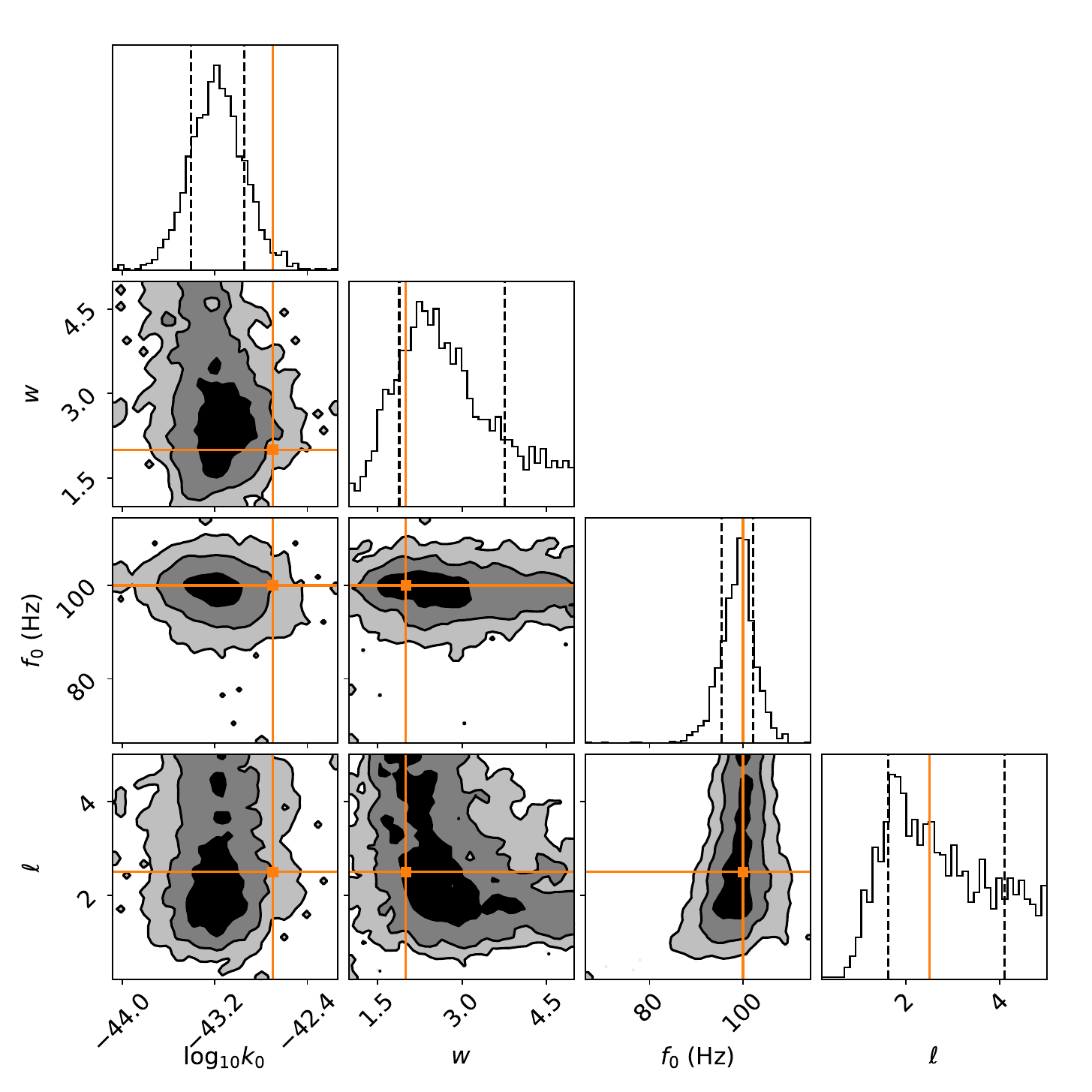}
    \caption{
    Posterior distribution of the Gaussian process hyper-parameters for a simulated signal with a deviation from GR drawn from our kernel. The contours are the $1$,$2$ and $3\sigma$ intervals. The true values are plotted as orange cross hairs, and are included within at least the $3\sigma$ interval for all hyper-parameters.
    }
    \label{fig:snr_5_corner}
\end{figure*}

In Fig. \ref{fig:snr_5_corner}, we show the posterior distribution for the Gaussian process hyper-parameters, with the true parameter values marked with orange cross hairs. 
This figure shows that our posteriors are consistent with the true hyper-parameter values.
The fact that the posterior for $\log_{10} k_0$ clearly excludes the minimum value of $k_0=10^{-46}$ illustrates that the algorithm is confident that a deviation from GR is present. 

\begin{figure*}
    \centering
    \includegraphics[width=\linewidth]{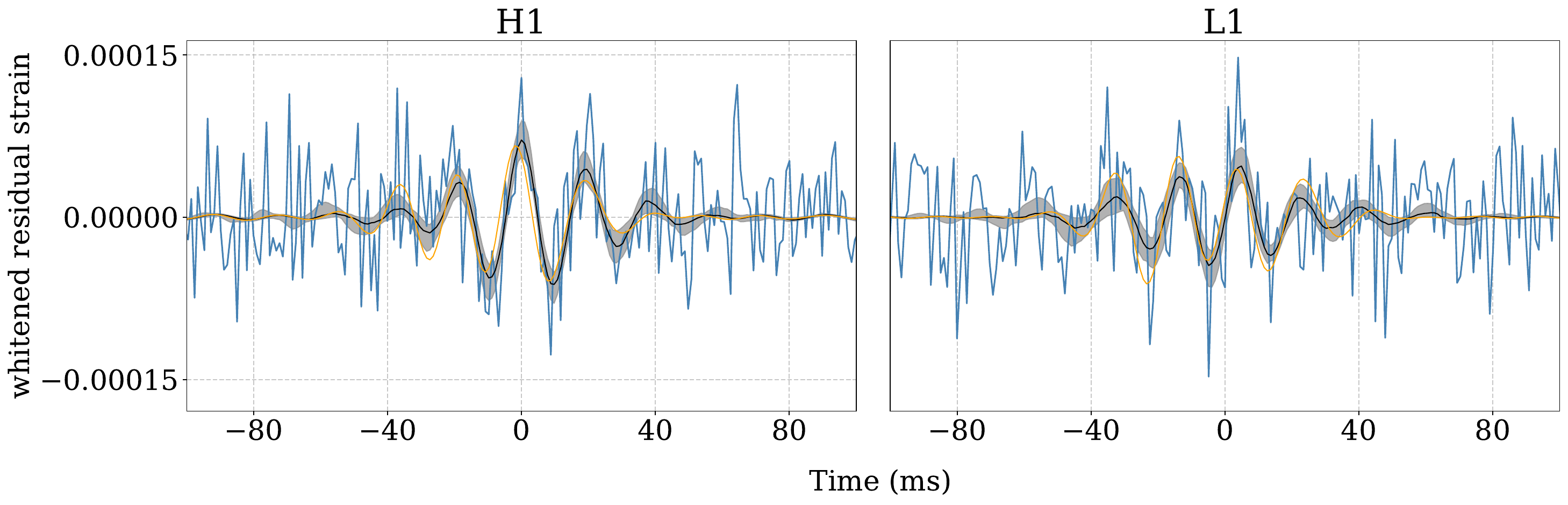}
    \caption{
    Reconstruction of the whitened deviation strain $\delta{s}$ for a simulated signal that includes a deviation from GR drawn from our kernel. The blue curve is the whitened strain in each detector. The orange curve is the true injected $\delta{s}$. The black curve is the median estimate of $\delta{s}$ using Gaussian process regression, and the grey shaded region is the $90\%$ credible interval on $\delta{s}$.
    }
    \label{fig:snr_5_reconstruction}
\end{figure*}

When a statistically significant signal is present, one can use the likelihood function to derive a posterior for $\delta s$ that marginalises over the Gaussian process hyper-parameters; see Appendix \ref{reconstruction} for details. 
We illustrate this process of signal reconstruction in Fig.~\ref{fig:snr_5_reconstruction}.
The reconstructed signal is consistent with the injected signal.

We define a signal-to-noise Bayes factor
\begin{equation}\label{bayesfactor}
\mathcal{B} = \frac{\int dk_0 \, \mathcal{L}(\delta h | k_0) \, \pi(k_0)}{\mathcal{L}(\delta h | k_0 = 0) \ },
\end{equation}
which is the ratio of the Bayesian evidence for $k_0>0$ to the Bayesian evidence for $k_0=0$, sometimes called the noise evidence.
If $\ln\mathcal{B} \ll 0$, the noise model is preferred over the signal model, and if $\ln{\cal B} \gg 0$, the signal model is preferred. 
Using Eq. \ref{bayesfactor}, we obtain $\ln{\mathcal{B}} = 21.9$ in support of the signal hypothesis for the optimal network SNR $=9.3$ signal injection, showing that we correctly find evidence for a deviation.

\subsection{Tests on noise}\label{noisesegments}
We next analyse segments of instrumental noise to check we do not confidently detect a signal when no such signal is present. 
We build a distribution of signal-to-noise Bayes factors (Eq. \ref{bayesfactor}), which allows us to convert a Bayes factor into a $p$-value.
This frequentist check is useful because real interferometer noise is non-Gaussian, and so a large Bayes factor might arise from misspecification of our noise model. For each binary black hole event from GWTC-3 that we analyse, we analyze three \unit[2]{s} long data segments from $\unit[110]s$, $\unit[120]s$ and $\unit[130]s$ before the event.
We make sure that both the LIGO H1 and L1 detectors are in observing mode and that no gravitational-wave signal has confidently been detected in these segments. 

We check that these data segments do not contain previously identified glitches by checking for the presence of either CAT1 or CAT2 data-quality vetoes \citep[see, e.g.][]{abbottOpenDataThird2023}. We also check for coincidence of any data segments with glitches identified by the \textsc{GravitySpy} framework \citep[][]{zevinGravitySpyIntegrating2017, glanzerDataQualityThird2023} with higher than 90\% confidence. We do not include a data segment if its center lies within \unit[1.5]{s} of such a glitch. 

To ensure the analysis of noise segments is as similar as possible to the analysis of real GW events, we inject a GW150914-like gravitational-wave signal into each segment. 
We perform parameter estimation to obtain posteriors on the binary parameters of the \textsc{IMRPhenomPv2} waveform. 
We use standard priors on the right ascension $\alpha$, declination $\delta$, inclination angle $\theta_{JN}$, azimuthal angle $\phi_{12}$, coalescence phase $\phi_c$, polarisation angle $\psi$, geocent time $t_0$, and spin tilts $\theta_1$ and $\theta_2$. We choose a uniform prior on the spin magnitudes $a_1$ and $a_2$ in the range $[0,0.99]$. We choose a uniform prior in comoving luminosity distance in the range $[10,10^4]\,\mathrm{Mpc}$. We constrain $m_1$ and $m_2$ in the range $[10,200]\,\mathrm{M_\odot}$, use a uniform prior in chirp mass $\mathcal{M}{}$ in the range $[20,60]\,\mathrm{M_\odot}$ and a uniform prior in mass ratio $q$ in the range $[0.05,1]$. We marginalise over time, distance and phase \citep[see, e.g.][]{thraneIntroductionBayesianInference2019} during sampling. 
We form residual data by subtracting the best-fit (maximum-likelihood) GR template from the data.
We perform Gaussian process regression on the remaining 174 signals using the procedure detailed in the preceding section.

\begin{figure}
    \centering
    \includegraphics[width=0.49\textwidth]{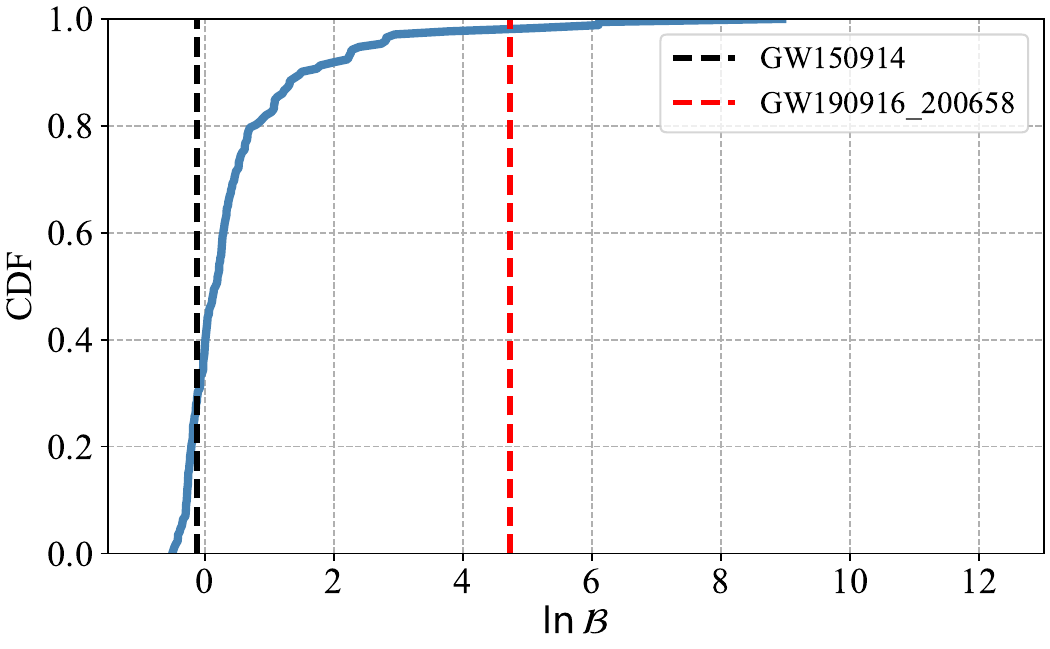}
    \caption{
    Cumulative distribution function (CDF) of the log-signal-to-noise Bayes factor $\ln{\mathcal{B}}$ for 174 segments of realistic noise, $\pi(\mathcal{B})$ (blue curve). Also indicated is the value of $\ln{\mathcal{B}}$ for GW150914 residual data (black dotted line) and for GW190916\_200658 residual data (red dotted line), the event with the highest $\ln{\mathcal{B}}$, though still typical of the distribution of $\ln{\mathcal{B}}$ for noise segments.
    }
    \label{fig:B_distribution}
\end{figure}

In Fig.~\ref{fig:B_distribution}, we show the cumulative distribution of $\ln{\mathcal{B}}$ for the 174 noise segments.
The black dashed line shows the value of $\ln{\cal B}$ for GW150914, and The red dashed line shows the value of $\ln{\cal B}$ for GW190916\_200658.
This figure shows that GW150914 residual data is broadly consistent with segments of realistic noise, and that the event with the most confident deviation from GR, GW190916\_200658, is still consistent with the these noise segments. Not shown in this plot is a single noise segment with $\ln{\mathcal{B}} \approx{200}$ containing what is likely a glitch that is not flagged by CAT1 or CAT2 data quality vetoes, and is not part of the \textsc{GravitySpy} glitch database. Though this noise segment is clearly an outlier with respect to the distribution of noise segments, we do not exclude it in our analysis.

\subsection{Tests with other signals}\label{db_2_injection}
To test the flexibility of our formalism, we analyse a GR-violating signal not drawn from our kernel.
In particular, we consider a GR-violating waveform where the signal is described by \textsc{IMRPhenomPv2} \citep[][]{schmidtModelsGravitationalWaveforms2012}, except the intermediate-stage post-Newtonian phase coefficient $\beta_2$ is changed from the value predicted by GR.
This parameter is particularly important for short-duration, binary black hole mergers \citep[][]{abbottTestsGeneralRelativity2016}.
Using the \textsc{Bilby TGR} package \citep[][]{ashton_bilby_2024}, we inject a gravitational-wave signal with GW150914-like parameters, but we change $\beta_2$ by $\delta \beta_2 = 1.0$.

We inject the gravitational-wave signal into $\unit[1]{s}$ of simulated Gaussian noise, coloured according to the LIGO design sensitivity. 
We perform parameter estimation to obtain posteriors on the binary parameters of the \textsc{IMRPhenomPv2} waveform. We use the same binary-parameter priors as in Section \ref{noisesegments}.
We form residual data by subtracting the best-fit (maximum-likelihood) GR template from the data. This residual data necessarily contains a deviation from GR (due to the imperfect subtraction of a GR signal from a non-GR signal), with an optimal network SNR $=12.1$. We center the data at the max-likelihood merger time of the event.
We perform Gaussian process regression on this segment, with the same procedure as before.

\begin{figure*}
    \centering
    \includegraphics[width=12cm]{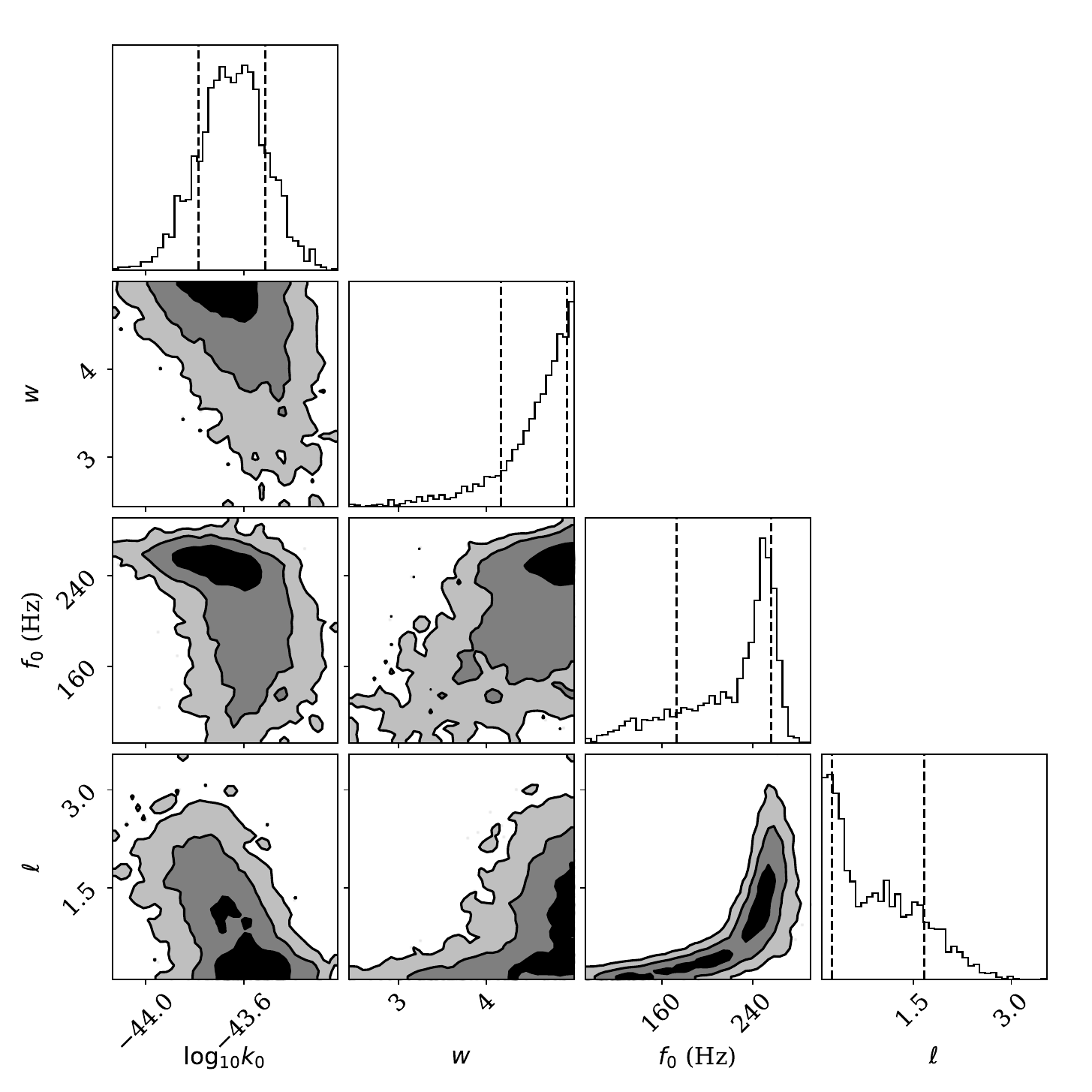}
    \caption{
    Posterior distribution of the Gaussian process hyper-parameters for a simulated gravitational-wave signal with a parameterised deviation from GR in the form of a $\delta \beta _2 \neq 0$ phase coefficient. The contours are the $1$,$2$ and $3\sigma$ intervals.}
    \label{fig:dbeta_2_corner}
\end{figure*}

\begin{figure*}
    \centering
    \includegraphics[width=\linewidth]{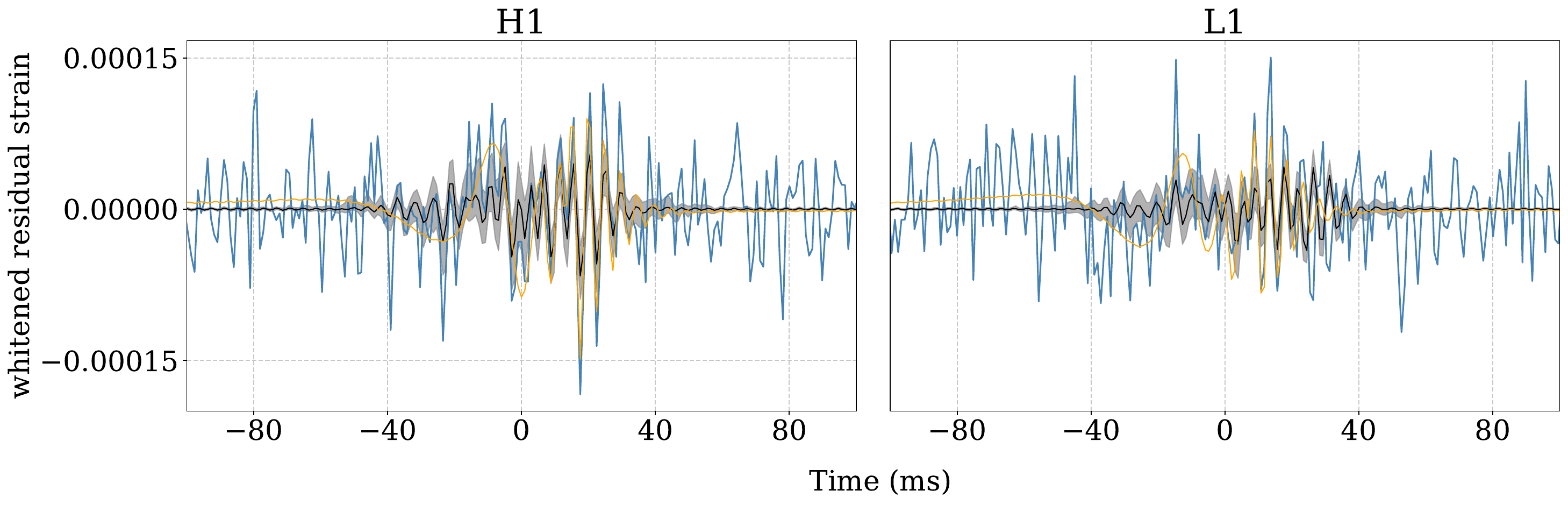}
    \caption{
    Reconstruction of the whitened deviation strain $\delta{s}$ for deviation from GR parameterised by the $\delta \beta_2$ phase coefficient. The blue curve is the whitened strain in each detector. The orange curve is the true injected $\delta{s}$. The black curve is the median estimate of $\delta{s}$ using Gaussian process regression, and the grey shaded region is the $90\%$ credible interval on $\delta{s}$.
    }
    \label{fig:dbeta_2_reconstruction}
\end{figure*}

In Fig. \ref{fig:dbeta_2_corner}, we show the posterior distribution for our kernel hyper-parameters.
We see that the minimum value of $k_0 = 10^{-46}$ is excluded from the $90\%$ credible interval. We plot the whitened deviation-strain reconstruction in Fig. \ref{fig:dbeta_2_reconstruction}, from which it is clear a deviation from GR is present. The signal-to-noise Bayes factor for this event is $\ln{\mathcal{B}} = 13.02$, with an associated $p$-value of $p=0.005$ when compared to the background distribution, suggesting strong statistical significance. 
The signal is better fit for times $t>0$, as this part of the signal is better represented using a single characteristic frequency, which matches the model, compared to the time-dependent frequency evolution for $t<0$. One could better model such `chirping' deviations using a kernel with multiple characteristic frequencies, or a time-dependent characteristic frequency. It also has higher frequency content than the signal at $t<0$, and since the noise at lower frequencies is greater, this part of the signal is higher SNR, and therefore better captured by the model.
This demonstrates our model is flexible enough to capture deviations from GR that are not drawn from the kernel.

\section{Results from GWTC-3}\label{residual_data}
We next apply our Gaussian process framework to the 60 binary black hole gravitational-wave events from GWTC-3 \citep[][]{ligoscientificcollaborationGWTC3CompactBinary2023} detected by both the H1 and L1 observatories. 
We do not consider any glitch-subtracted data frames, instead using the publicly available GWOSC data \citep[][]{abbottOpenDataThird2023}.
We perform parameter estimation to obtain posteriors on the binary parameters of the \textsc{IMRPhenomPv2} waveform for \unit[2]{s} of data around each event, using the same approach as in subsections \ref{noisesegments} and \ref{db_2_injection}. We produce residual data for each event by subtracting the maximum-likelihood estimate \textsc{IMRPhenomPv2} waveform, and center the data at the maximum-likelihood merger time of each event.
We perform Gaussian process regression on each of these data segments, as in the previous sections. We use the same priors on the Gaussian process hyper-parameters as outlined in subsection \ref{simulateddeviations}.

\begin{figure*}[ht]
    \centering
    \includegraphics[width=12cm]{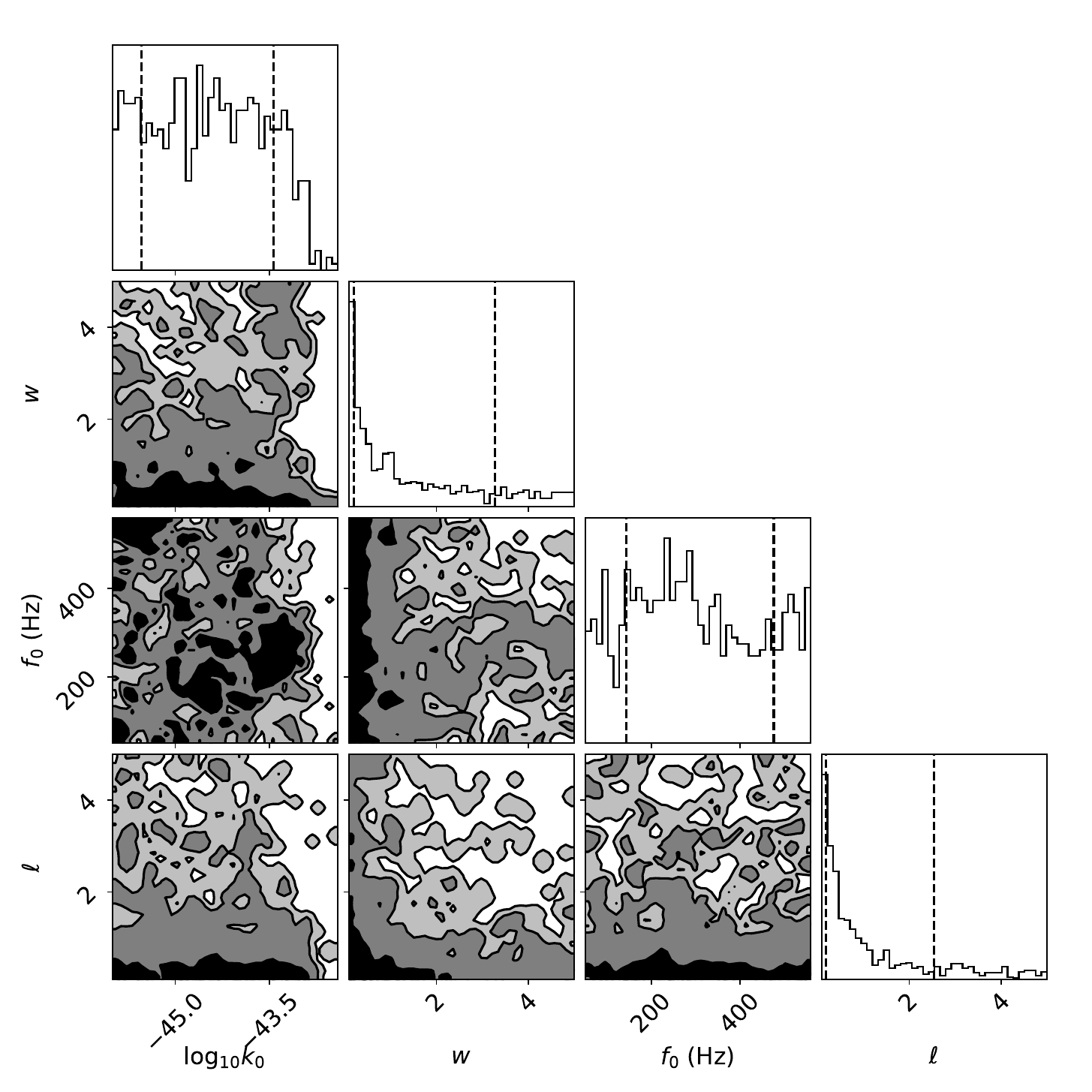}
    \caption{
    Posterior distribution of the Gaussian process hyper-parameters for GW150914 residual data. The contours are the $1$,$2$ and $3\sigma$ intervals. The posterior on $k_0$ does not exclude the lower bound of $10^{-46}$, indicating no signal is present in the data.
    }
    \label{fig:GW150914_corner}
\end{figure*}

\begin{figure*}
    \centering
    \includegraphics[width=\linewidth]{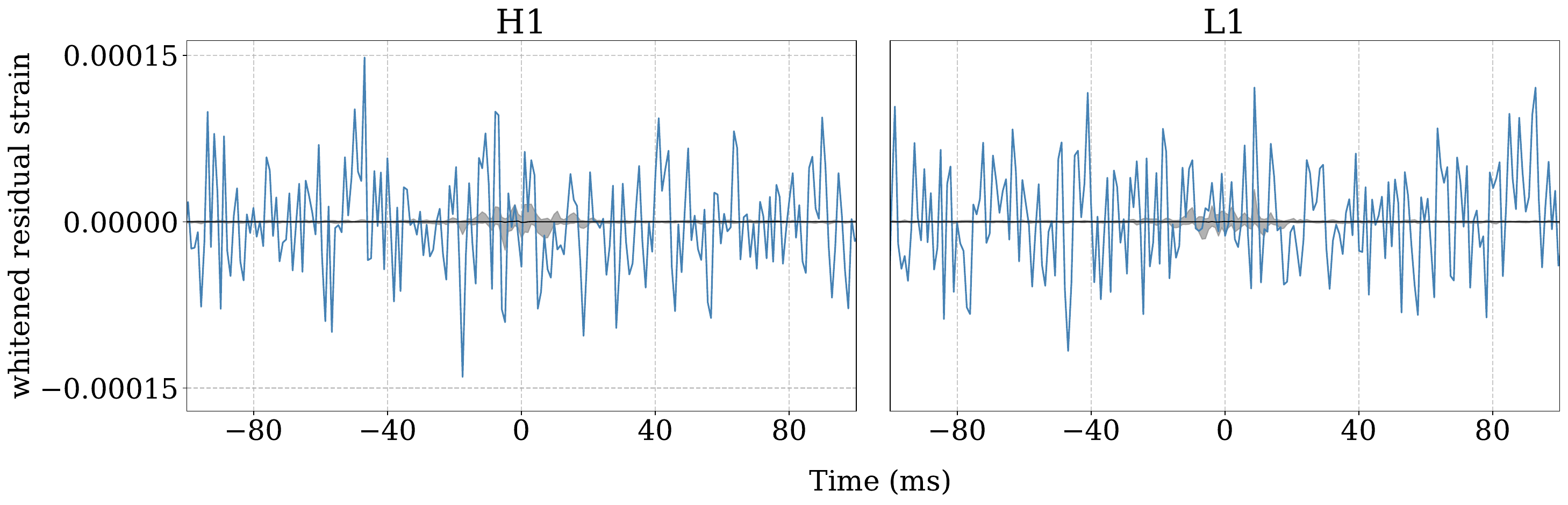}
    \caption{
    Reconstruction of the whitened deviation strain $\delta{s}$ for GW150914 residual data. The blue curve is the whitened residual strain in each detector. The black curve is the median estimate of $\delta{s}$ using Gaussian process regression, and the grey shaded region is the $90\%$ credible interval on $\delta{s}$, which shows no signal is present in the data.
    }
    \label{fig:GW150914_reconstruction}
\end{figure*}

As an illustrative example, we show the Gaussian process hyper-posteriors for GW150914 in Fig.~\ref{fig:GW150914_corner}. For this event, we find that the lower bound of our $k_0$ prior, $k_0 = 10^{-46}$, is included in the $90\%$ credible interval, indicating that a deviation from GR is unlikely. In Fig. \ref{fig:GW150914_reconstruction}, we show the reconstruction of $\delta s$, which confirms the absence of a deviation from GR. Equation \ref{bayesfactor} yields a Bayes factor of $\ln{\mathcal{B}} = -0.12$, corresponding to a $p$-value of $0.69$, suggesting a deviation from GR is not present in GW150914 residual data. 

\begin{figure}
    \centering
    \includegraphics[width=0.49\textwidth]{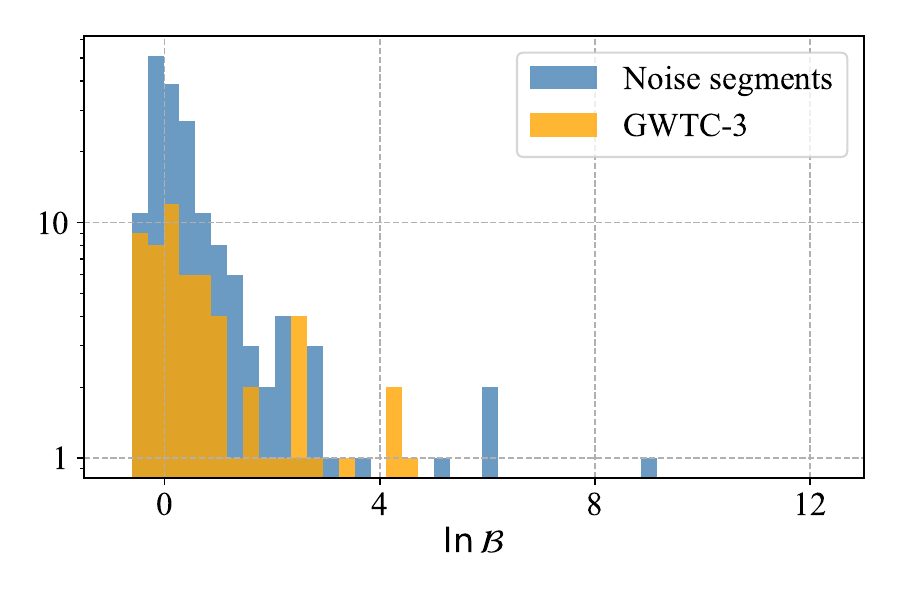}
    \caption{
    Comparison between the log-signal-to-noise Bayes factor $\ln{\mathcal{B}}$ distributions for 174 noise segments (blue) and the 60 GWTC-3 GW events (orange) analysed. 
    }
    \label{fig:comparison_ln_B}
\end{figure}

In Fig. \ref{fig:comparison_ln_B} we show the distribution of $\ln{\mathcal{B}}$ for all the GW events that we analyse.
We find that it is consistent with the distribution of $\ln{\mathcal{B}}$ from 174 noise segments analysed previously. We show the signal-to-noise Bayes factor for all individual events in Appendix \ref{GWTC3_table}. 
No events show evidence for a deviation from GR.
If any events were not consistent with the distribution of $\ln{\mathcal{B}}$ for 174 noise segments, it is possible that this number of noise segments is too small. We would increase the number of noise segments to be sure we are not misrepresenting this distribution.
We find that the event with the largest $\ln{\mathcal{B}}$ is GW190916\_200658, with $\ln{\mathcal{B}} = 4.71$ ($p=0.03$), which is unsurprising given that we looked at 60 events, so we expect $p$-values ${\cal O}(1/60) \approx 2\%$ even if the data are well described by GR.

We note that there is one noise outlier in the distribution with $\ln {\cal B}\approx 200$.
The presence of such outliers harms the sensitivity of the search because only very loud signals will produce deviations from GR with larger Bayes factors.
It may be possible to remove these outliers through an aggressive data quality investigation.
We leave this for future work.

Since the data are consistent with GR, we calculate upper limits on the fractional strain of the non-GR deviation $\delta{s}$ for each GW event:
\begin{enumerate}
    \item We take random draws of hyper-posterior samples $\Lambda_k$.
    \item For each sample, we take a random draw of $\delta s(t)$, the time-domain deviation strain.
    \item We record the maximum value of 
    \begin{align}
        \delta s_\text{max} = \max_t \left|\delta s(t)\right| , \nonumber
    \end{align}
    in order to obtain a posterior distribution on $\delta s_\text{max}$.
    \item We find the 90\% credible interval for $\delta s_\text{max}$.
    \item We take random draws of waveform posterior samples $\theta_i$.
    \item For each sample, we calculate the quadrature sum of the maximum $+$ and $\times$ polarisation strain amplitudes for the \textsc{IMRPhenomPv2} waveform, $\delta s_{GR,\text{max}}$.
    \item We find the 90\% credible interval for $\delta s_{GR,\text{max}}$.
    \item We calculate the fractional strain as ${\delta s_\text{max}}/{\delta s_{\text{GR,max}}}$
\end{enumerate}
We constrain the maximum fractional deviation-strain amplitude to be as low as $7\%$ (for GW190701\_203306).

As the residual data for each event is constructed using a maximum-likelihood waveform estimate, one may worry about deviations arising from imperfect subtraction of the GR template. 
Fortunately, our analysis can be extended to marginalise over uncertainty in the GR waveform.
As simultaneous inference of both the gravitational waveform and Gaussian process hyper-parameters is computationally expensive, importance sampling can be used to reweight the result of our initial analysis---obtained with the maximum-likelihood template---to the result we would have obtained by marginalising over the gravitational waveform parameters. 

We detail this process in Appendix \ref{marginalising}. 
We demonstrate this for the event with the most significant deviation only, GW190916\_200658, and find the significance shifts from $\ln{\mathcal{B}} = 4.71$ to $\ln{\mathcal{B}} =3.85$. As this change is relatively minor, and since marginalisation over the uncertainty in the GR template tends to decrease the significance of potential deviations from GR, we do not perform this marginalisation for the other events studied. 

\section{Discussion}\label{discussionconclusions}
When searching for a deviation from GR in gravitational-wave residual data, in the absence of a single compelling alternate theory, one should adopt a flexible model to cast a broad net. To this end, we introduce a Gaussian process formalism to search for deviations from GR in gravitational-wave residual data. We encode our prior beliefs in our choice of kernel. We apply this formalism to the 60 gravitational-wave events from GWTC-3 detected by both the H1 and L1 gravitational-wave observatories and find no evidence of a deviation from GR, and constrain the maximum fractional deviation strain amplitude to as low as $7\%$ for GW190701\_203306. 

If a deviation from GR is detected with the framework introduced in this work, there are two explanations. The first is that the signal detected is truly a deviation from GR. The second---far more likely explanation---is that the models used for analysis are \textit{misspecified}; \citep{wmf,guptaPossibleCausesFalse2024}. For example, the noise model could be misspecified; e.g., non-Gaussian noise may be present in the data. 
To show that signal is not due to noise-model misspecification, one can inject simulated GR signals into off-source noise in order to build up an empirical distribution of Bayes factors (as we demonstrate above).
This distribution can be used to calculate an empirical $p$-value, which should include the effects of unmodeled noise. 

The signal model may also be misspecified. For example, our use of the \textsc{IMRPhenomPv2} waveform may not make accurate assumptions for all events we analyse (e.g. that they all have aligned spins). Therefore, in the case one finds a deviation from GR, more state-of-the-art waveforms should be used. 
Even in the case that such a waveform is used, it is extremely challenging to distinguish between misspecification in the waveform-approximant and a deviation from GR --- all waveforms have some inherent assumptions of the underlying physics. 
A natural question arises: what conclusions can be drawn from residual tests of GR if we cannot confidently differentiate between a true deviation from GR and signal-model misspecification? 

Our position is that the method presented here is not enough. 
This framework merely provides a tool for detecting deviations from our best waveform approximants.
In the event that a deviation is detected, modelers would need to determine if the deviation could plausibly be explained by systematic error in the waveform approximant. 
Repeated deviations from the best template banks, which---after a thorough investigation---cannot be plausibly explained as systematic errors, would force us to consider seriously the alternative hypothesis: that we have detected a deviation from GR.

Our Gaussian-process framework is computationally expensive, particularly for longer-duration gravitational-wave signals, which require a larger covariance matrix to accommodate the increased frequency resolution. The likelihood evaluation requires inversion of the covariance matrix (see Eq. \ref{eq:L}), an operation that scales as $\mathcal{O}(n^3)$, where $n$ is the number of dimensions of the covariance matrix. 
In the future, we hope to investigate methods such as the \textsc{celerite} package \citep[][]{foreman-mackeyFastScalableGaussian2017,ashtonGaussianProcessesGlitchrobust2023}, which take advantage of specific forms of the covariance matrix to improve the scaling relation to $\mathcal{O}(n \log{n})$.

Similar to \citet[][]{ashtonGaussianProcessesGlitchrobust2023}, it may be possible to extend this framework to model noise glitches in gravitational-wave data. This would require revision of the kernel used in analysis, to either loosen the coherency constraint between detectors, or introduce new kernel functions/hyperparameters to more suitably model typical glitches. For example, one could allow for additional characteristic frequencies, allow for time-dependent (`chirping') characteristic frequencies, or allow the peak of the signal to be different to the merger time. We leave this as an avenue for future research.

\section*{Acknowledgements}
This is LIGO document \#P2500394.
We acknowledge support from the Australian Research
Council (ARC) Centres of Excellence CE170100004 and CE230100016, as well as ARC
LE210100002, and ARC DP230103088. L.P. and S.Y.C.
receive support from the Australian Government Research
Training Program. 
This material is based upon work supported by NSF’s LIGO Laboratory
which is a major facility fully funded by the National
Science Foundation. The authors are grateful for computational resources provided by the LIGO Laboratory
and supported by National Science Foundation Grants
PHY-0757058 and PHY-0823459.

This research has made use of data or software obtained from the Gravitational Wave Open Science Center (gw-openscience.org), a service of LIGO Laboratory,
the LIGO Scientific Collaboration, the Virgo Collaboration, and KAGRA. LIGO Laboratory and Advanced
LIGO are funded by the United States National Science Foundation (NSF) as well as the Science and Technology Facilities Council (STFC) of the United Kingdom, the Max-Planck-Society (MPS), and the State of
Niedersachsen/Germany for support of the construction
of Advanced LIGO and construction and operation of
the GEO600 detector. Additional support for Advanced
LIGO was provided by the Australian Research Council.
Virgo is funded, through the European Gravitational
Observatory (EGO), by the French Centre National
de Recherche Scientifique (CNRS), the Italian Istituto
Nazionale di Fisica Nucleare (INFN) and the Dutch
Nikhef, with contributions by institutions from Belgium,
Germany, Greece, Hungary, Ireland, Japan, Monaco,
Poland, Portugal, Spain. The construction and operation of KAGRA are funded by Ministry of Education,
Culture, Sports, Science and Technology (MEXT), and
Japan Society for the Promotion of Science (JSPS), National Research Foundation (NRF) and Ministry of Science and ICT (MSIT) in Korea, Academia Sinica (AS)
and the Ministry of Science and Technology (MoST) in
Taiwan.

\section*{Data Availability}

The data underlying this article are publicly available at \url{https://www.gw-openscience.org}.

\bibliographystyle{aasjournal}
\bibliography{refs}

\appendix
\section{Covariance matrices for different numbers of detectors}\label{different_detectors}
While we use only the strain data from two detectors in our analysis, it is simple to extend our framework to different numbers of detectors. For a single detector $\mu$, only self-covariances exist:

\begin{align}\label{eq:Cij2}
    \textbf{\textit{C}}_{ij}^{\mu} = & 
    \left(
    \begin{matrix}
        (F_{\mu, +}^2 + F_{\mu,\times}^2) \textbf{\textit{K}}_{ij} + \textbf{\textit{N}}_{ij}^\mu
    \end{matrix}
    \right) .
\end{align}

For three detectors $\mu$, $\nu$ and $\rho$, the rank of $\textbf{\textit{C}}_{ij}$ increases, and appropriate covariances between detectors must be considered:

\begin{align}\label{eq:Cij}
    \textbf{\textit{C}}_{ij}^{\mu\nu\rho} = & 
    \left(
    \begin{matrix}
        (F_{\mu, +}^2 + F_{\mu,\times}^2) \textbf{\textit{K}}_{ij} + \textbf{\textit{N}}_{ij}^\mu &  e^{2\pi i f_j \tau_{\mu\nu}} (F_{\mu,+}F_{\nu,+} + F_{\mu,\times}F_{\nu,\times}) \textbf{\textit{K}}_{ij} &  
        e^{2\pi i f_j \tau_{\mu\rho}} (F_{\mu,+}F_{\rho,+} + F_{\mu,\times}F_{\rho,\times}) \textbf{\textit{K}}_{ij}\\
         e^{-2\pi i f_j \tau_{\mu\nu}} (F_{\mu,+}F_{\nu,+} + F_{\mu,\times}F_{\nu,\times}) \textbf{\textit{K}}_{ji}^*  & (F_{\nu,+}^2 + F_{\nu,\times}^2) \textbf{\textit{K}}_{ij} + \textbf{\textit{N}}_{ij}^\nu & e^{2\pi i f_j \tau_{\nu\rho}} (F_{\nu,+}F_{\rho,+} + F_{\nu,\times}F_{\rho,\times}) \textbf{\textit{K}}_{ij}  \\
         e^{-2\pi i f_j \tau_{\mu\rho}} (F_{\mu,+}F_{\rho,+} + F_{\mu,\times}F_{\rho,\times}) \textbf{\textit{K}}_{ji}^* &
         e^{-2\pi i f_j \tau_{\nu\rho}} (F_{\nu,+}F_{\rho,+} + F_{\nu,\times}F_{\rho,\times}) \textbf{\textit{K}}_{ji}^* &
         (F_{\rho,+}^2 + F_{\rho,\times}^2) \textbf{\textit{K}}_{ij} + \textbf{\textit{N}}_{ij}^\rho
    \end{matrix}
    \right) .
\end{align}
A similar treatment may be applied in the case of four or more detectors.

\section{Reconstructing the deviation}\label{reconstruction}
We wish to reconstruct the deviation from GR $\delta s$ given residual data $\delta h$ -- that is, we want to construct ${p}(\delta s | \delta h)$. We first expand Eq. \ref{eq:L} using Eqs. \ref{eq:C} and \ref{eq:Sij}, giving
\begin{align} \label{eq:B3}
    {\cal L}(\delta h |\delta s) \propto \frac{1}{ \text{det}(\textbf{\textit{N}})} \exp \Big( -( \delta h- \delta s)^\dagger \textbf{\textit{N}}^{-1} (\delta h- \delta s) \Big) 
    \, 
    \frac{1}{\text{det}(\textbf{\textit{S}}(\Lambda))}
    \exp \Big( - \delta s^\dagger \textbf{\textit{S}}^{-1}(\Lambda)  \delta s \Big) .
\end{align}    

The posterior on $\delta s$ given $\delta h$ and hyper-parameters $\Lambda$ is given by Bayes theorem
\begin{align}
    {p}( \delta s | \delta h, \Lambda) = &
    \frac{{\cal L} (\delta h | \delta s) \pi (\delta s | \Lambda) \pi (\Lambda)}{\cal Z}, 
\end{align}
\text{where marginalising over $\Lambda$ gives}
\begin{align}
    {p}( \delta s | \delta h) = &
    \int d\Lambda \,
    p( \delta s | \delta h, \Lambda) \\
    = &
    \int d\Lambda \,
    \frac{{\cal L} (\delta h | \delta s) \pi (\delta s | \Lambda) \pi (\Lambda)}{\cal Z}.
\end{align}

\text{Multiplying by unity gives}
\begin{align}
    p( \delta s | \delta h)
    = &
    \int d\Lambda \,
    \frac{{\cal L} (\delta h | \delta s) \pi (\delta s | \Lambda) \pi (\Lambda)}{\cal Z} \frac{{p}(\Lambda | \delta h)}{{p}(\Lambda | \delta h)},
\end{align}
\text{which we simplify and approximate as a sum over posterior samples $k$ of $\Lambda$} 
\begin{align}
    p( s | h)
    \propto & \sum_k
    \frac{{\cal L} (\delta h | \delta s) \pi (\delta s | \Lambda_k) }{{\cal L} (\Lambda_k | \delta h)}. \\
\end{align}

Expanding this using Eq. \ref{eq:B3} gives
\begin{align}
    p( \delta s | \delta h)
    \propto & \sum_k
    \frac{1}{\text{det}(\textbf{\textit{S}}(\Lambda_k))}
    \frac{1}{\text{det}(\textbf{\textit{N}})}
    \frac{1}{{\cal L}(\Lambda_k | \delta h)}
    \exp \Big( -(\delta h-\delta s)^\dagger \textbf{\textit{N}}^{-1} (\delta h-\delta s) \Big) 
    \exp \Big( -\delta s^\dagger \textbf{\textit{S}}^{-1}(\Lambda_k) \delta s \Big). 
\end{align}

We now ``complete the square'', giving
\begin{align}
    p( \delta s | \delta h)
    \propto & \sum_k
    \exp \Big( - \frac{1}{2}(\delta s-(\textbf{\textit{N}}^{-1} + \textbf{\textit{S}}^{-1}(\Lambda_k))^{-1}\textbf{\textit{N}}^{-1}\delta h)^\dagger (\textbf{\textit{N}}^{-1}+\textbf{\textit{S}}^{-1}(\Lambda_k)) (\delta s-(\textbf{\textit{N}}^{-1}+\textbf{\textit{S}}^{-1}(\Lambda_k))^{-1}\textbf{\textit{N}}^{-1}\delta h) \Big) \notag\\
    & \quad \times \exp \Big( - \frac{1}{2} \delta h^\dagger \textbf{\textit{N}}^{-1} (\textbf{\textit{N}}^{-1} \textbf{\textit{S}}^{-1}(\Lambda_k))^{-1}\textbf{\textit{N}}^{-1}-I)\delta h \Big), 
\end{align}
which simplifies to
\begin{align}
    p( \delta s | \delta h)
    \propto & \sum_k
    \exp \Big( - \frac{1}{2}(\delta s-(\textbf{\textit{N}}^{-1} + \textbf{\textit{S}}^{-1}(\Lambda_k))^{-1}\textbf{\textit{N}}^{-1}\delta h)^\dagger (\textbf{\textit{N}}^{-1}+\textbf{\textit{S}}^{-1}(\Lambda_k)) (\delta s-(\textbf{\textit{N}}^{-1}+\textbf{\textit{S}}^{-1}(\Lambda_k))^{-1}\textbf{\textit{N}}^{-1}\delta h) \Big).
\end{align}

This is a multivariate Gaussian distribution with mean and variance
\begin{align}
    \mu = (\textbf{\textit{N}}^{-1} + \textbf{\textit{S}}^{-1}(\Lambda_k))^{-1}\textbf{\textit{N}}^{-1}\delta h ,\\
    \Sigma = (\textbf{\textit{N}}^{-1}+\textbf{\textit{S}}^{-1}(\Lambda_k))^{-1} .
\end{align}

\section{Marginalising over uncertainty in the gravitational waveform}\label{marginalising}
Our analysis uses only the max-likelihood estimate gravitational waveform parameters to produce residual data. We can account for uncertainty in the waveform parameters by marginalising over them using importance sampling. Our target likelihood is the likelihood of observing $d$, given $\Lambda$, marginalised over all $\theta$,
\begin{align}
   {\cal L} (d | \Lambda) = \int  d \theta {\cal L} (d | \Lambda, \theta)  \pi(\theta)  .
\end{align}
To perform importance sampling, we desire a relation between the unknown likelihood ${\cal L} (d | \Lambda)$ and the known posterior samples $p (\theta | d)$. To achieve this, we first multiply by unity:
\begin{align}
   {\cal L} (d | \Lambda) = \int d \theta {\cal L} (d | \Lambda, \theta)  \pi(\theta)   \frac{p(\theta|d)}{p(\theta|d)}.
\end{align}
We can now write this as a sum over posterior samples $\theta_i$:
\begin{align}
   {\cal L} (d | \Lambda) \approx \frac{1}{N_\theta} \sum_{i=1}^{N_\theta} 
   \frac{{\cal L} (d| \Lambda, \theta_i)\pi(\theta_i)}
   {p(\theta_i|d)}.
\end{align}
Applying Bayes theorem gives 
\begin{align}\label{eq:waveform_reweighting}
   {\cal L} (d | \Lambda) \approx \frac{\mathcal{Z}_\theta}{N_\theta} \sum_{i=1}^{N_\theta} 
   \frac{{\cal L} (d| \Lambda, \theta_i)}
   {{\cal L}(d|\theta_i)},
\end{align}
where we have canceled the factor $\pi(\theta_i)$ in the numerator and denominator. Here, $\mathcal{Z}_\theta$ is the evidence for the gravitational waveform model. We desire the reweighted evidence $\mathcal{Z}_\text{RW,GP}$ that takes into account waveform uncertainty:
\begin{align}
   \mathcal{Z}_\text{RW,GP} = \int {d \Lambda {\cal L} (d | \Lambda) \pi(\Lambda)} \approx \int d \Lambda \frac{\mathcal{Z}_\theta}{N_\theta} \sum_{i=1}^{N_\theta} 
   \frac{{\cal L} (d| \Lambda, \theta_i)}
   {{\cal L}(d|\theta_i)} \pi(\Lambda).
\end{align}
To utilise the posterior $p(\Lambda | d, \theta_\text{ML})$ we again multiply by unity:
\begin{align}
   \mathcal{Z}_\text{RW,GP} \approx \frac{\mathcal{Z}_\theta}{N_\theta} \int d\Lambda \sum_{i=1}^{N_\theta} 
   \frac{{\cal L} (d| \Lambda, \theta_i)}
   {{\cal L}(d|\theta_i)} \frac{p(\Lambda | d, \theta_\text{ML})}{p(\Lambda | d, \theta_\text{ML})}\pi(\Lambda).
\end{align}
We now write our expression as a double sum over posterior samples $\theta_i$ and $\Lambda_k$:
\begin{align}
   \mathcal{Z}_\text{RW,GP} \approx 
   \frac{\mathcal{Z}_\theta}{N_\theta N_\Lambda} \sum_{i,k=1}^{N_\theta, N_\Lambda} 
   \frac{{\cal L} (d| \Lambda_k, \theta_i)}
   {{\cal L}(d|\theta_i)} \frac{\pi(\Lambda_k)}{p(\Lambda_k | d, \theta_\text{ML})}.
\end{align}
We once again use Bayes theorem:
\begin{align}
   \mathcal{Z}_\text{RW,GP} \approx 
   \frac{\mathcal{Z}_\theta \mathcal{Z}_\Lambda}{N_\theta N_\Lambda} \sum_{i,k=1}^{N_\theta, N_\Lambda} 
   \frac{{\cal L} (d| \Lambda_k, \theta_i)}
   {{\cal L}(d|\theta_i){\cal L}(d | \Lambda_k, \theta_\text{ML})} ,
\end{align}
where we have canceled the factor $\pi(\Lambda_k)$ in the numerator and denominator. Here, $\mathcal{Z}_\Lambda$ is the evidence for the Gaussian process analysis that used the max-likelihood estimate waveform parameters $\theta_\text{ML}$. We are interested in the reweighted Bayes factor that takes into account waveform uncertainty,
\begin{align}
    \mathcal{B}_\text{RW} = \frac{\mathcal{Z}_\text{RW,GP}}{\mathcal{Z}_\text{RW,Noise}}.
\end{align}
$\mathcal{Z}_\text{RW,Noise}$ is simply given by Equation \ref{eq:waveform_reweighting},
\begin{align}
   \mathcal{Z}_\text{RW,Noise} = {\cal L} (d | \Lambda(k_0=0) \approx \frac{\mathcal{Z}_\theta}{N_\theta} \sum_{i=1}^{N_\theta} 
   \frac{{\cal L} (d| \Lambda(k_0=0), \theta_i)}
   {{\cal L}(d|\theta_i)}.
\end{align}
Substituting into the equation for $\mathcal{B}_\text{RW}$ and simplifying gives
\[
\mathcal{B}_{\mathrm{RW}}
\approx
\frac{%
  \displaystyle
  \frac{\mathcal{Z}_{\Lambda}}{N_{\Lambda}}
  \sum_{i=1}^{N_{\theta}}\sum_{k=1}^{N_{\Lambda}}
  \dfrac{\mathcal{L}\bigl(d\!\mid\!\Lambda_{k},\theta_{i}\bigr)}
        {\mathcal{L}\bigl(d\!\mid\!\theta_{i}\bigr)\,
         \mathcal{L}\bigl(d\!\mid\!\Lambda_{k},\theta_{\mathrm{ML}}\bigr)}
}{%
  \displaystyle
  \sum_{j=1}^{N_{\theta}}
  \dfrac{\mathcal{L}\bigl(d\!\mid\!\Lambda(k_0 = 0),\theta_{j}\bigr)}
        {\mathcal{L}\bigl(d\!\mid\!\theta_{j}\bigr)}
}
\]

The weight corresponding to each $\Lambda_k$ is then
\begin{align}
    w_k \equiv \sum_{i=1}^{N_\theta} \frac{{\cal L} (d | \Lambda_k, \theta_i)}{{\cal L} (d | \theta_i) {\cal L} (d | \Lambda_k, \theta_\text{ML})},
\end{align}

When reweighting, one should take care that the number of effective samples is suitably large \citep[see, e.g.,][]{hom}, by calculating the quantity
\begin{align}
    n_\text{eff} = \frac{(\sum_k{w_k})^2}{\sum_k w_k^2} .
\end{align}

\begin{figure*}[ht]
    \centering
    \includegraphics[width=12cm]{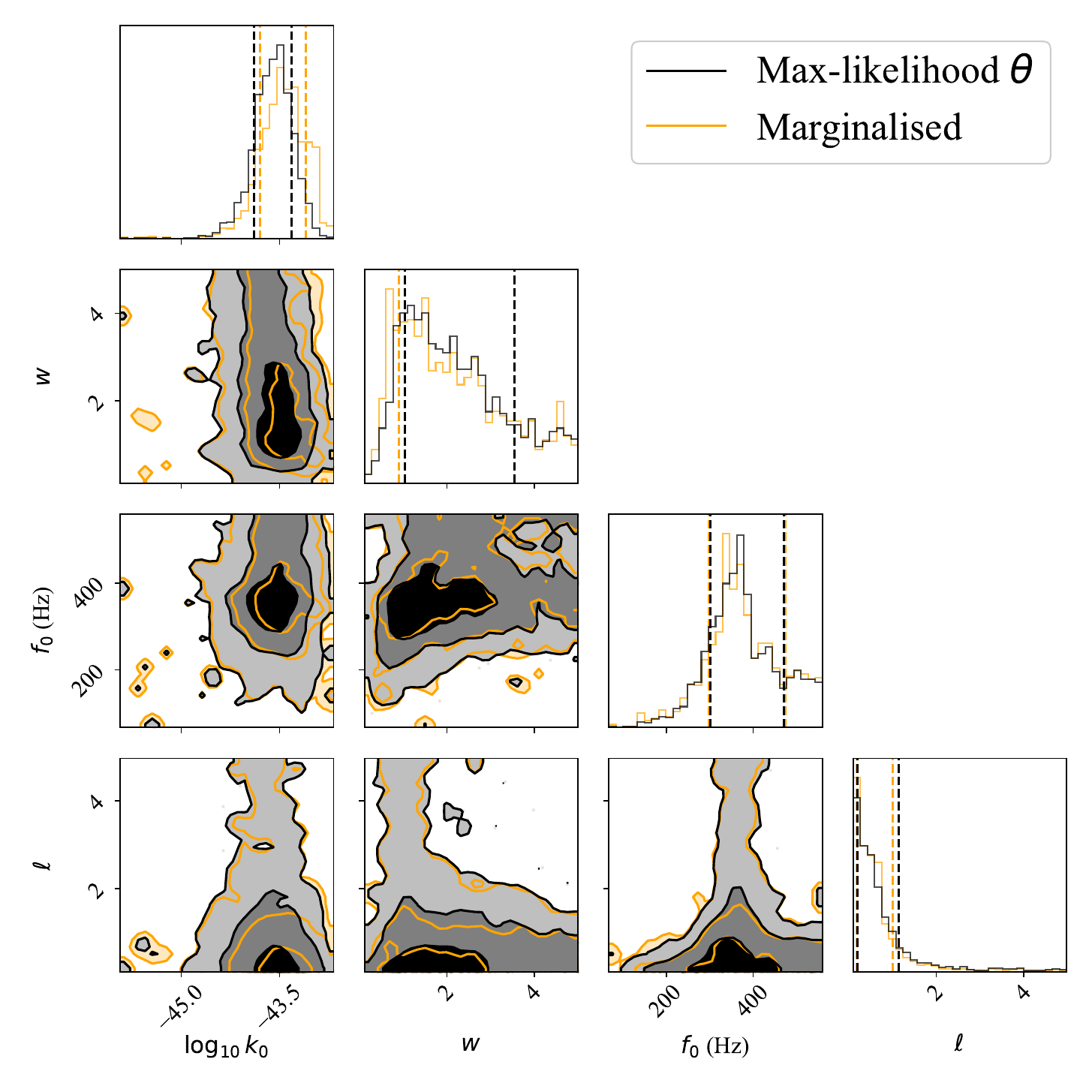}
    \caption{
    Posterior distribution of the Gaussian process hyper-parameters for GW190916\_200658 residual data. The contours are the $1$,$2$ and $3\sigma$ intervals. Shown in black are the posteriors for the initial max-likelihood estimate run for the GR waveform parameters $\theta$, and shown in orange are the re-weighted posteriors, taking into account the uncertainty in $\theta$.}
    \label{fig:marginalisation_corner}
\end{figure*}

\begin{figure*}
    \centering
    \includegraphics[width=\linewidth]{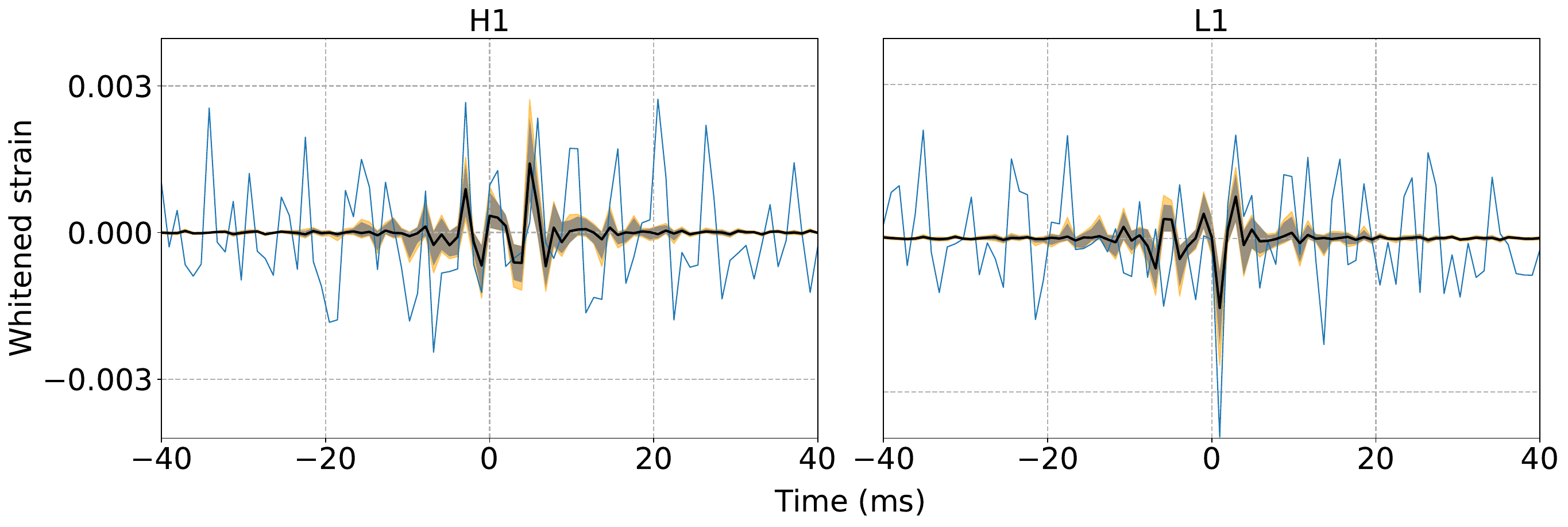}
    \caption{
    Reconstruction of the whitened deviation strain $\delta{s}$ for GW190916\_200658 residual data. The blue curve is the whitened residual strain in each detector. The black curve is the median estimate of $\delta{s}$ using Gaussian process regression for the max-likelihood estimate of the GR waveform parameters $\theta$, and the grey shaded region is the $90\%$ credible interval on $\delta{s}$ for this estimate. The orange shaded region is the $90\%$ credible interval on $\delta{s}$ marginalised over the uncertainty in $\theta$.}
    \label{fig:marginalisation_reconstruction}
\end{figure*}

We now demonstrate this for the event with the highest significance,  GW190916\_200658. We reweight 1500 randomly chosen hyper-posterior samples of $\Lambda$ to account for marginalising over the uncertainty in our GR waveform model (which is itself sampled by another set of 1000 posterior draws). We show the effects of this on both the hyper-parameter posteriors and the reconstructed deviation in Figures \ref{fig:marginalisation_corner} and \ref{fig:marginalisation_reconstruction}. We find the reconstructed deviation broadens minimally, and its significance shifts from $\ln{\mathcal{B}} = 4.71$ to $\ln{\mathcal{B}} =3.85$, which we believe is relatively minor, and thus do not perform this treatment for other events. The number of hyper-posterior samples is $1500$, and $n_\text{eff} = 356$, indicating a reweighting efficiency of $24\%$, which we believe is sufficient for this demonstration.

\newpage
\section{Results for GWTC-3}\label{GWTC3_table}
\begin{table}[h!]

\centering
\caption{Summary of the $\ln{\mathcal{B}}$ for all GWTC-3 gravitational wave events analysed. A value of $\ln{\mathcal{B}} > 0$ indicates support for the hypothesis that the data are not fully described by the \textsc{IMRPhenomPv2} waveform.}
\begin{tabular}{|c|c|}
\hline
\textbf{Event} & \(\ln\mathcal{B}\) \\
\hline
GW150914            & -0.12 \\
GW151012            &  0.07 \\
GW170104            &  0.35 \\
GW170729            & -0.02 \\
GW170809            & -0.22 \\
GW170814            &  0.18 \\
GW170818            &  0.90 \\
GW170823            &  2.79 \\
GW190403\_051519    &  1.94 \\
GW190408\_181802    &  0.80 \\
GW190412            &  0.93 \\
GW190413\_052954    &  1.65 \\
GW190413\_134308    &  0.60 \\
GW190421\_213856    &  0.13 \\
GW190426\_190642    &  0.53 \\
GW190503\_185404    &  0.24 \\
GW190512\_180714    & -0.34 \\
GW190513\_205428    &  0.27 \\
GW190514\_065416    & -0.21 \\
GW190517\_055101    &  2.59 \\
GW190519\_153544    &  2.54 \\
GW190521            &  3.43 \\
GW190521\_074359    &  0.71 \\
GW190527\_092055    & -0.24 \\
GW190602\_175927    &  0.63 \\
GW190701\_203306    &  0.07 \\
GW190706\_222641    &  0.49 \\
GW190719\_215514    &  1.52 \\
GW190727\_060333    &  0.53 \\
GW190731\_140936    &  2.06 \\
\hline
\end{tabular}
\quad
\begin{tabular}{|c|c|}
\hline
\textbf{Event} & \(\ln\mathcal{B}\) \\
\hline
GW190803\_022701    &  0.03 \\
GW190805\_211137    &  0.85 \\
GW190828\_063405    & -0.40 \\
GW190828\_065509    &  4.37 \\
GW190915\_235702    &  0.07 \\
GW190916\_200658    &  4.71 \\
GW190926\_050336    & -0.43 \\
GW190929\_012149    &  1.12 \\
GW191109\_010717    &  0.27 \\
GW191113\_071753    & -0.45 \\
GW191127\_050227    &  1.08 \\
GW191204\_110529    & -0.32 \\
GW191215\_223052    & -0.32 \\
GW191222\_033537    & -0.08 \\
GW191230\_180458    &  1.17 \\
GW200128\_022011    &  0.35 \\
GW200129\_065458    &  0.76 \\
GW200208\_130117    &  4.28 \\
GW200208\_222617    & -0.56 \\
GW200209\_085452    &  2.43 \\
GW200216\_220804    & -0.32 \\
GW200219\_094415    & -0.05 \\
GW200220\_061928    & -0.01 \\
GW200220\_124850    &  2.59 \\
GW200224\_222234    & -0.32 \\
GW200225\_060421    &  0.08 \\
GW200306\_093714    &  0.45 \\
GW200308\_173609    &  0.03 \\
GW200311\_115853    & -0.08 \\
\hline
\end{tabular}

\label{tab:bf_table}
\end{table}

\end{document}